\documentclass[a4paper]{jpconf}
\usepackage{graphicx}
\begin{document}
\title{Phase Transitions in the Two-Dimensional Ising Model 
from the Microcanonical Perspective}
\author{Kedkanok Sitarachu and Michael Bachmann}
\address{Soft Matter Systems Research Group, Center for Simulational Physics, 
Department of Physics and Astronomy, University of Georgia, Athens, GA 30602, 
USA}
\ead{Kedkanok.Sitarachu@uga.edu, bachmann@smsyslab.org\\ \textrm{Homepage:} 
http://www.smsyslab.org}
\begin{abstract}
The continuous ferromagnetic-paramagnetic phase transition in the 
two-dimensional Ising model has already been excessively studied by 
conventional 
canonical statistical analysis in the past. We use the recently developed 
generalized microcanonical 
inflection-point analysis method to investigate the least-sensitive 
inflection 
points of the microcanonical entropy and its derivatives to identify 
transition signals. Surprisingly, this method reveals that there are 
potentially two additional
transitions for the Ising system besides the critical transition.   
\end{abstract}
\section{Introduction}
The Lenz-Ising model~\cite{lenz1,ising1} for spin systems provides an 
excellent foundation for the study of long-range correlations in a system 
with short-range interactions only. In the two-dimensional 
(2D) case, this cooperative behavior of local spins leads to a thermodynamic 
phase transition between the disordered paramagnetic and the ordered 
ferromagnetic phase~\cite{onsager1}. However, the one-dimensional (1D) 
model does not 
show significant signs of cooperativity~\cite{ising1}. 
Although the 1D and 2D models were solved exactly many decades ago, the Ising 
model has inspired a multitude of studies thereafter and is probably the 
best-studied system exhibiting complex behavior in statistical physics. 

Almost all of these studies have focused on canonical statistical 
properties of the critical behavior of the 2D system in the thermodynamic 
limit. Here, we follow a different conceptual approach and use recently 
expanded ideas of microcanonical thermodynamics~\cite{gross1,Bachmann2014} 
for the analysis. The generalized microcanonical inflection-point analysis 
method~\cite{qb1} was developed to incorporate systems of 
finite size in the theory of phase transitions and to introduce a robust 
classification system for phase transitions that does not require the 
extrapolation toward the thermodynamic limit. This is particularly important 
for systems that are naturally finite, such as biological macromolecules, but 
the method can be used for the study of phase behavior in any system of 
interest, independently of system size.

As we will show in the following, microcanonical inflection-point analysis 
does not only correctly signal the expected second-order phase transition 
between the ferromagnetic and the paramagnetic phase. It also reveals two 
additional transitions of higher order, which do not seem to disappear in 
the thermodynamic limit. 
\section{Microcanonical analysis of least-sensitive inflection points}
The typical classification of phase transitions is based on Ehrenfest's 
scheme, in which the order of a transition is determined by the order of the 
derivative of the appropriate thermodynamic potential (typically the Gibbs 
free enthalpy) with respect to its natural variables that exhibits a 
discontinuity at the transition point in the thermodynamic 
limit~\cite{ehrenfest1}. It thus makes use of the macroscopic state variables 
of the system, which are represented by averages over microscopic degrees of 
freedom in the canonical statistical ensemble. Since virtually all 
thermodynamic phase transitions that fit into this scheme are first- or 
second-order phase transitions, it is nowadays more common to distinguish 
only discontinuous and continuous transitions, respectively.

A significant drawback of the conventional canonical statistical analysis is 
that a unique 
identification of a phase transition relies on the necessity to study the 
system in the thermodynamic limit or to extrapolate to it in order to 
create a non-analyticity that signals a phase transition. This is not a 
problem for very large systems, for which the scaling to infinite size is 
possible and reasonable. 
However, for finite systems, there are no discontinuities. Therefore, in 
studies of 
systems like proteins, whose sizes cannot be extrapolated toward the 
thermodynamic limit, it has become popular to locate ``peaks'' and 
``shoulders'' in canonical response quantities such as the 
specific heat, susceptibility, and fluctuations of order parameters. One 
problem with this approach is that these signals are not unique; the 
location of a transition point typically depends on the quantity used for its 
identification (see, e.g., Fig.~5 in Ref.~\cite{bj1}). 

Another issue is that 
a peak in a fluctuating quantity does not necessarily signal a transition at 
all. The most prominent example is the 1D Ising model. There is no phase 
transition, but the specific heat curve possesses a pronounced peak, which 
remains finite even in the thermodynamic limit. Thus, if a system does not 
allow for the extrapolation toward the thermodynamic limit, a peak may or may 
not signal a transition. Yet, cooperative processes like the folding of a 
protein have many features in common with phase transitions. Therefore, it 
would be useful to have an analysis method that can provide unique transition 
signals even for systems of finite size. Microcanonical inflection-point 
analysis has been developed for this purpose and, in addition, allows for a 
systematic, hierarchical classification of the transitions~\cite{qb1}.

The microcanonical entropy
\begin{equation}
S(E)=k_\mathrm{B}\ln\, g(E),
\end{equation}
where $g(E)$ is the density (or number) of states with energy $E$, and its 
derivatives $\beta(E)=dS(E)/E$, $\gamma(E)=d^2S(E)/dE^2$, 
$\delta(E)=d^3S(E)/dE^3$, $\dots$ have a well-defined monotony in energy 
regions that do not include transition points. 
Thus, changes in monotony can be considered signals of 
peculiar behavior. Since entropy and energy drive thermodynamic phase 
transitions, these monotonic changes~-- represented by inflection points 
in the curves of $S(E)$ or its derivatives~-- are indicative signals
of transitions in the system. More specifically, since transitions are 
typically associated with large fluctuations in energy, 
least-sensitive inflection points in these curves can serve as signals for 
transitions. The order of the transition is in correspondence with
the order of the derivative of $S$ that exhibits a 
least-sensitive inflection point~\cite{qb1}. For example, a least-sensitive 
inflection point in the first derivative, $\beta(E)$, indicates a 
second-order transition.

It is important to note that a least-sensitive inflection point can (but 
does not necessarily have to) occur together with a satellite signal at 
higher energy, which, if it is present, is always of higher order. This makes 
it necessary to distinguish \emph{independent} and \emph{dependent} 
transitions~\cite{qb1}.
In the generalized inflection-point analysis method, an \emph{independent 
transition} of odd order $(2k-1)$ ($k$ 
positive integer) is characterized by
\begin{equation}
\left. \frac{d^{(2k-1)}S(E)}{dE^{(2k-1)}} 
\right|_{E=E_{\mathrm{tr}}}>0, 
\label{eq:odd_order_trans}
\end{equation}
whereas for even order $2k$
\begin{equation}
\left. \frac{d^{2k}S(E)}{dE^{2k}} \right|_{E=E_{\mathrm{tr}}}<0. 
\label{eq:even_order_trans}
\end{equation}
Since the dependent transitions are always of higher order than the 
transition they correspond to, their minimal order is 2. 
\emph{Dependent transitions} of even order $2k$ satisfy
\begin{equation}    
\left. \frac{d^{2k}S(E)}{dE^{2k}} \right|_{E=E_\mathrm{tr}^\mathrm{dep}}>0, 
\label{eq:even_order_dep_trans}
\end{equation}
and for odd-order transitions
\begin{equation}
\left. \frac{d^{(2k+1)}S(E)}{dE^{(2k+1)}}
\right|_{E=E_\mathrm{tr}^\mathrm{dep}}<0. 
\label{eq:odd_order_dep_trans}
\end{equation}
In the following, we apply this method to the 2D Ising model.
\section{Results}
The energy of a configuration $\mathbf{S}=(s_1,s_2,\ldots,s_{L^2})$ of $L^2$ 
locally interacting spins with possible states $s_i=\pm 1$ on the $L\times L$ 
square lattice 
with periodic boundary conditions (i.e., a torus) is expressed in the 
Lenz-Ising model~\cite{lenz1,ising1} by
\begin{equation}
E(\mathbf{S})=-J\sum\limits_{\langle i,j\rangle} s_is_j,
\end{equation}
where $\langle i,j\rangle$ symbolizes that the sum only runs over pairs of 
spins that are nearest neighbors on the lattice. If $J>0$, neighboring spins 
energetically prefer a parallel orientation, which can cause spontaneous 
ordering under appropriate thermodynamic conditions. Hence, the model allows 
to 
describe the phase transition between the paramagnetic phase of disordered 
spin configurations and the ferromagnetic phase with nonzero spontaneous 
magnetization. For antiferromagnetic coupling, $J<0$, the ordered phase is 
dominated by configurations with an alternating spin pattern. In the 
following, for simplicity, 
we only discuss cases with even numbers of spins $L$ in each row and column. 
This renders the density of states symmetric in energy space. In this case, 
the choice of the $J$ sign does not actually matter and the results are 
identical for ferromagnetic and antiferromagnetic systems under the symmetry 
$E\to -E$. For convenience, we choose $J\equiv 1$ (and also set
$k_\mathrm{B}\equiv 1$) in the following. The microcanonical inverse 
temperature $\beta$ is negative for the non-equilibrium states with $E>0$ 
in the Ising model. Since we here intend to focus on the equilibrium phases, 
any signals in the positive-valued $E$ space will be ignored. 
\begin{figure}
\centerline{\includegraphics[width=16cm]{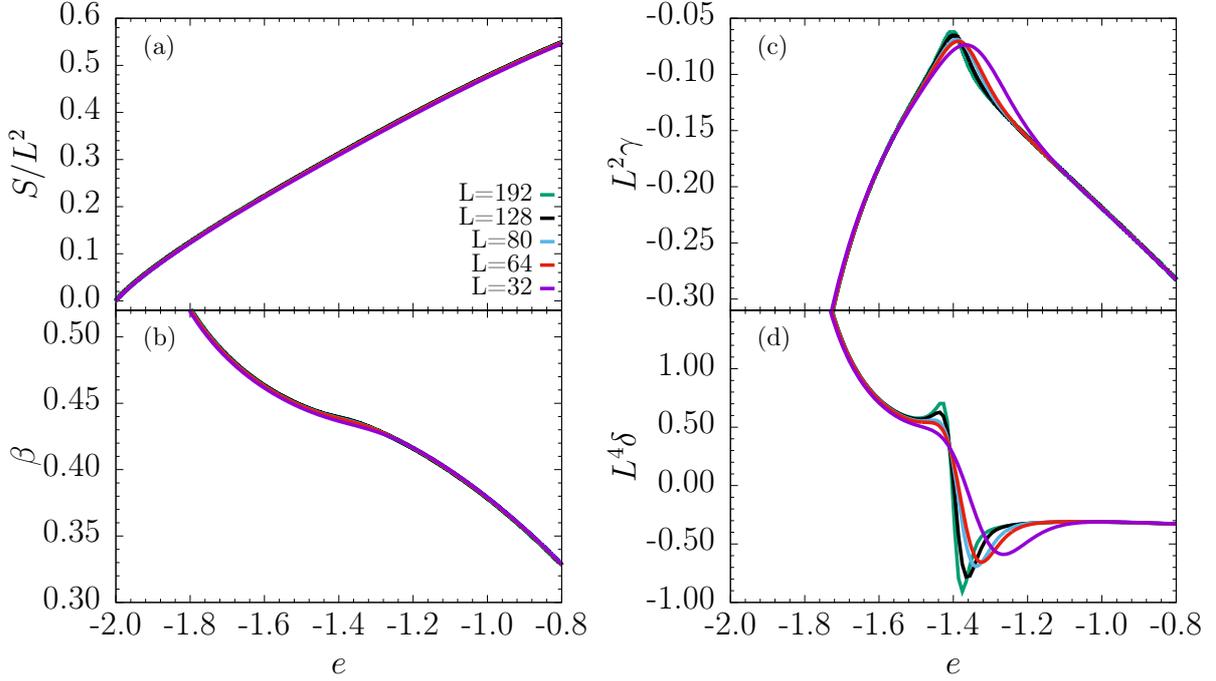}}
\caption{\label{fig:micro2D}
Microcanonical entropy per spin $S/L^2$, $\beta$, $L^2\gamma$, and 
$L^4\delta$ plotted as functions of the energy per spin, $e=E/L^2$ for 
various system sizes.}
\end{figure}
\begin{figure}
\centerline{\includegraphics[width=10cm]{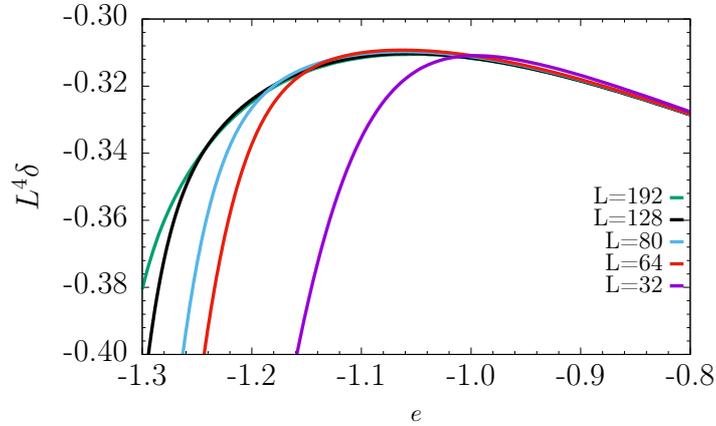}}
\caption{\label{fig:delta}
Enlarging the section of $L^4\delta$ above the critical transition in 
Fig.~\ref{fig:micro2D}(d) makes the peaks associated with an additional 
third-order 
dependent transition clearly visible.}
\end{figure}

For the analysis of the properties of the microcanonical entropy and its 
derivatives, we employ Beale's algorithmic method~\cite{beale1} for the exact 
evaluation of the density of states. The derivatives are calculated 
numerically using symmetric difference expressions.
The thus obtained exact microcanonical results for systems with up to $192^2$ 
spins are shown in Figs.~\ref{fig:micro2D} and~\ref{fig:delta}. 

We immediately recognize that, to leading order, the entropy $S$ scales with 
system size [$\sim\mathcal{O}(L^2)$] in the space of the re-scaled energy 
$e=E/L^2$ [Fig.~\ref{fig:micro2D}(a)]. The microcanonical inverse 
temperature $\beta$ is virtually scale independent in the same space 
[Fig.~\ref{fig:micro2D}(b)]. Since the entropy curves do not exhibit any 
least-sensitive inflection point (and, consequently, there is no extremum in 
$\beta$), a first-order transition does not exist.

The higher-order derivatives $\gamma$ and 
$\delta$, shown in Figs.~\ref{fig:micro2D}(c) and~(d), respectively, exhibit 
a nontrivial dependence on the system size within a narrow energy region. 
Outside this region, $\gamma\sim\mathcal{O}(1/L^2)$ and 
$\delta\sim\mathcal{O}(1/L^4)$ in the reduced energy space. The dependence 
and direction of change of these quantities with system size in the energy 
interval $e\in[-1.6,-1.0]$ is important. It determines if transitions in this 
energy region survive in the thermodynamic limit. 

\begin{figure}
\centerline{\includegraphics[width=10cm]{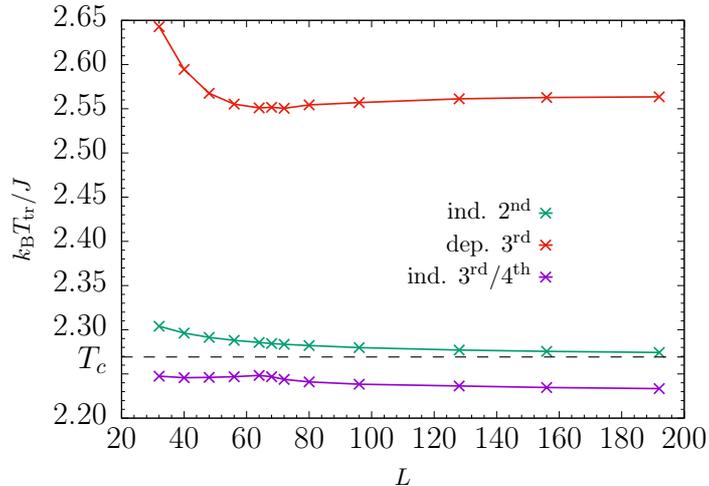}}
\caption{\label{fig:TvsL}
Transition temperatures $T_\mathrm{tr}$ as obtained by microcanonical 
analysis plotted as a function of $L$. Symbols indicate the system sizes, for 
which the actual calculations were performed. Lines connecting these points 
are guides to the eye. The curves do not suggest the merging of the 
three transition lines. 
We thus conclude that the additional higher-order transitions above and 
below the second-order ferromagnetic-paramagnetic transition line (which 
converges to $k_\mathrm{B}T_c/J=2/\ln(1+\sqrt{2})$ as expected; dashed line) 
could survive even in the 
thermodynamic limit. Note the change in curvature of the upper-critical 
transition line near $L\approx 64$, where the transition 
characteristics changes from fourth to third order.}
\end{figure}

First of all, we clearly notice the peaks in the $\gamma$ curves, associated 
with the respective least-sensitive inflection points in the $\beta$ curves. 
According to our classification scheme, this signal indicates a second-order 
transition. For the system sizes studied, we also observe that the peak in 
$\gamma$ becomes higher and sharper with increasing system size. It eventually 
will converge to $\gamma=0$ in the thermodynamic limit $L\to\infty$ as this is 
the familiar second-order phase transition between the ferromagnetic and 
paramagnetic phase. The position of the $\gamma$ peaks in $e$ space and the 
corresponding $\beta$ values at the inflection point of the individual 
systems enable the estimation of the transition temperature 
$T_\mathrm{tr}(L)$ for each of the finite systems. In Fig.~\ref{fig:TvsL}, we 
plot the dependence of the transition temperature on $L$. The points 
associated with the second-order transition clearly converge toward the exact 
critical transition temperature $T_c=2/\ln(1+\sqrt{2})$. This 
confirms that microcanonical inflection-point analysis identifies the known 
second-order phase transition qualitatively and quantitatively correctly, as 
expected.

What is surprising, however, is the fact that the minimum that develops in 
$\delta$ with increasing system size at energies (and 
temperatures) below the critical 
point indicates an independent third-order transition (for $L \le 64$ it is 
of fourth order) according 
to our classification scheme. Likewise, above the critical point, there is 
another signal in $\delta$. The peak near $e\approx -1.1$ is interpreted as 
a dependent third-order transition, though, which is associated with the 
critical transition. Although the relative peak height on these scales does 
not increase with system size [Fig.~\ref{fig:delta}], there is no clear 
sign of its extinction either (in fact the nearby minimum becomes deeper and 
deeper) and from the data available we conclude this transition might also 
exist in the thermodynamic limit. 

The dependence of the 
corresponding transition temperatures on the system size is also shown in 
Fig.~\ref{fig:TvsL}. For the system sizes studied, the exact results do not 
suggest a merger of these additional transitions with the critical transition 
in the thermodynamic limit. Microcanonical analysis cannot hint at the nature 
of these transitions; therefore an in-depth structural analysis of the 
results obtained in stochastic computer simulations for larger systems is 
necessary in future work.
\section{Summmary}
We have employed the generalized microcanonical inflection-point analysis 
method~\cite{qb1} for the study of the transition behavior of the 2D Ising 
model on a square lattice
with periodic boundary conditions. The analysis was based on the exact 
densities of states obtained by using Beale's evaluation method~\cite{beale1} 
for various system sizes with up to $192^2$ spins. The method correctly 
identifies the known second-order phase transition and the transition points 
obtained for the finite systems converge to the exactly known critical 
temperature in the thermodynamic limit. In addition, inflection-point analysis 
reveals two additional transitions, which are of higher-than-second order. 
One is located in the subcritical regime and it exists independently of the 
critical transition. For systems with $L > 64$, it is of third order. For 
smaller systems it shows the characteristics of a fourth-order transition. 
The other transition, which is of third order and found above the critical 
point, would not exist without the critical transition and thus depends on it. 
The origin and microscopic features of these additional transitions are not 
known yet. Based on the available results, no indication of a convergence of 
the transition points toward the critical temperature in the thermodynamic 
limit was found, which leads to the conclusion that they might also exist in 
the thermodynamic limit.
\section*{Acknowledgments}
We thank Dr.\ Kai Qi for helpful information.

\section*{References}
\begin{thebibliography}{99}
%
\bibitem{lenz1}
Lenz W 1920 Beitrag zum Verst\"andnis der magnetischen Erscheinungen in 
festen K\"orpern \textit{Z.~Phys.} \textbf{21} 613
%
\bibitem{ising1}
Ising E 1925 Beitrag zur Theorie des Ferromagnetismus \textit{Z.~Phys.} 
\textbf{31} 253
%
%
\bibitem{onsager1}
Onsager L 1944 Crystal Statistics. I. A Two-Dimensional Model with an 
Order-Disorder Transition \textit{Phys.\ Rev.} \textbf{65} 117
%
\bibitem{gross1}
Gross D H E 2001 \textit{Microcanonical Thermodynamics} (Singapore: World 
Scientific)
%
\bibitem{Bachmann2014}
Bachmann M 2014 \textit{Thermodynamics and Statistical Mechanics
of Macromolecular Systems} (Cambridge: Cambridge University Press)
%
\bibitem{qb1}
Qi K and Bachmann M 2018 Classification of Phase Transitions by 
Microcanonical 
Inflection-Point Analysis \textit{Phys.\ Rev.\ Lett.} \textbf{120} 180601
%
\bibitem{ehrenfest1}
Ehrenfest P 1933 Phasenumwandlungen im ueblichen und erweiterten Sinn,
classifiziert nach den entsprechenden Singularitaeten des thermodynamischen
Potentiales \textit{Proc.\ Royal Acad.\ Amsterdam (Netherlands)} \textbf{36} 
153; \textit{Commun.\ Kamerlingh Onnes Inst. Leiden} Suppl.\ No.\ 75b
%
\bibitem{bj1}
Bachmann M and Janke W 2004 Thermodynamics of Lattice Polymers 
\textit{J.~Chem.\ Phys.} \textbf{120} 6779
%
\bibitem{beale1}
Beale P D 1996 Exact Distribution of Energies in the Two-Dimensional Ising 
Model \textit{Phys.\ Rev.\ Lett.} \textbf{76} 78
%



%
\end{thebibliography}
\end{document}